\documentclass[aps,preprint,epsfig,rotate]{revtex4}
\usepackage{graphicx}
\usepackage{bm}
\usepackage{epsfig}

\begin{document}
\title{Ionization of light atoms and ions during the nuclear $\beta^{-}$-decay}

\author{Alexei M. Frolov}
\email[E--mail address: ]{afrolov@uwo.ca}

\affiliation{Department of Applied Mathematics \\
 University of Western Ontario, London, Ontario N6H 5B7, Canada}

\author{David M. Wardlaw}
 \email[E--mail address: ]{dwardlaw@mun.ca}

\affiliation{Department of Chemistry, Memorial University of Newfoundland, St.John's, 
             Newfoundland and Labrador, A1C 5S7, Canada}

\date{\today}

\begin{abstract}

Ionization of light atoms and ions during the nuclear $\beta^{-}$-decay is considered. To determine the final state probabilities of electron 
ionization we have developed a procedure based on the natural orbital expansions for the bound state wave functions of all atoms/ions involved 
in this process.

\end{abstract}

\maketitle
\newpage

\section{Introduction}

In our earlier studies (see, e.g., \cite{Talm}, \cite{Our1}, \cite{Our2} and references therein) we considered atomic excitations during the 
nuclear $\beta^{-}$-decay in light, few-electron atoms and ions. In general, the $\beta^{-}$ decay of an atom is written in the form 
\begin{equation}
  X \rightarrow Y^{+} + e^{-}(\beta) + \overline{\nu} \label{equa1}
\end{equation}
where the symbols $X$ and $Y$ designate two different chemical elements (isotopes) with equal (or almost equal) masses. The sybmols X and Y in 
Eq.(\ref{equa1}) are used to designate the both atoms/ions and the corresponding atomic nuclei. If $Q$ is the electric charge of the incident 
nucleus $X$, then the nuclear charge of the final nucleus $Y$ is $Q + 1$. In Eq.(\ref{equa1}) and everywhere below the notation $e^{-}(\beta)$ 
stands for the fast electron (or $\beta^{-}$-electron) emitted from the nucleus during the nuclear $\beta-$decay, while $\overline{\nu}$ 
designates the electron's anti-neutrino. In \cite{Talm} - \cite{Our2} we assumed that the incident atom was in one of its bound state and the 
final ion is also formed in one of its bound states. This means that in \cite{Talm} - \cite{Our2} we discussed the bound-bound transitions between 
the incident and final states in few-electron atoms and/or ions. In the case of the decay, Eq.(\ref{equa1}), the incident atom and final ion 
contain the same numbers of bound electrons and this fact substantially simplifies numerical computations of the final state probabilities. In 
particular, in \cite{Our2} we have found that the nuclear $\beta^{-}$-decay of the ${}^{8}$Li and ${}^{9}$Li atoms in $\approx$ 85 \% of all cases 
lead to the formation of the three-electron ${}^{8}$Be$^{+}$ and ${}^{9}$Be$^{+}$ ions, i.e.  
\begin{equation}
 {}^{N}{\rm Li} \rightarrow {}^{N}{\rm Be}^{+} + e^{-}(\beta) + \overline{\nu} \label{eq1}
\end{equation}
in a variety of bound states. In Eq.(\ref{eq1}) the symbol $N$ means the total number of nucleons in the nuclei (in our case $N$ = 8, 9). The final 
state probabilities, i.e. probabilities to form different bound states in the Be$^{+}$ ion during the nuclear $\beta^{-}$-decay of the three-electron 
${}^{8}$Li and ${}^{9}$Li atoms have been accurately evaluated in \cite{Our2}.

Here we consider other atomic processes which occur during the nuclear $\beta^{-}$-decay in atoms and ions. The most important of such processes is the 
`additional' electron ionization of the final ion \cite{Mig}, \cite{MigK}. The general equation for this process takes the form:
\begin{equation}
  X \rightarrow Y^{2+} + e^{-} + e^{-}(\beta) + \overline{\nu} \label{equa2}
\end{equation}
where the notation $e^{-}$ stands for the slow (atomic) electron formed in the unbound spectrum during the reaction, Eq.(\ref{eq2}), while the notation 
$e^{-}(\beta)$ designates the fast $\beta^{-}$ electron. Note that after the process, Eq.(\ref{equa2}), the final $Y^{2+}$ ion has larger electric charge (+2), 
than the analogous $Y^{+}$ ion formed in the reaction, Eq.(\ref{equa1}). In the case of $\beta^{-}$-decay of the ${}^{8}$Li and ${}^{9}$Li atoms such an 
`additional' ionization of the three-electron Be$^{+}$ ion has $\approx$ 15 \% probability. This particular process is written in the form
\begin{equation}
 {\rm Li} \rightarrow {\rm Be}^{2+} + e^{-} + e^{-}(\beta) + \overline{\nu} \label{eq2}
\end{equation}
In some earlier works on $\beta^{-}-$decay in many-electron atoms the emited atomic electrons were called and considered as the `secondary' electrons, or 
$\delta-$electrons. The analogous process of `additional' electron ionization with the emission of the secondary $\delta-$electrons may proceed in any atom/ion 
where nuclear excitations are re-distributed between the atomic nucleus and bound electrons. The corresponding probabilities of an `additional' ionization of the 
Be$^{+}$ ion can be measured and used for a complete description of atomic excitations during the nuclear $\beta^{\pm}$-decay in the three-electron Li atom.  
The energy spectrum of the emitted $\delta-$electrons is another (unique) characteristic of the atomic $\beta^{-}$ decay. Further analysis shows that the secondary 
electron from the process, Eq.(\ref{eq2}), can be detected in different spin states. In an ideal case we can observe such an electron either in the $\alpha-$spin 
state, or in the $\beta-$spin state, where the notations $\alpha$ and $\beta$ are used to designate spin-up and spin-down wave functions, respectively (see, e.g., 
\cite{LLQ}). Since the total electron spin of the incident atom is conserved during the nuclear $\beta^{\pm}-$decay in a few-electron atom (see below), we can 
conclude that the final Be$^{2+}$ ion arising after the reaction, Eq.(\ref{eq2}), can be found in one of its singlet $S_e = 0$, or triplet $S_e = 1$ states. Here 
the notation $S_e$ stands for the total electron spin. In general, the final ion arising during the nuclear $\beta^{\pm}-$decay of the neutral atom is always 
formed in one of the two possible spin states and the difference of the spin values for such states equals unity, i.e. $\Delta S_e = 1$.

In reality, such newly formed states atomic are often unstable and this means double ionization of the final ion arising during the nuclear $\beta^{\pm}$-decay. 
Indeed, consider, for instance, the $\beta^{-}-$decay of some neutral atom which has four electrons in its outer-most electron shells. Ionization during this nuclear 
$\beta^{-}$-decay leads to the formation of an ion with three electrons in its outer-most shells. The electron configuration of such an ion corresponds to either 
doublet spin-state (e.g., ${}^{2}S-$state), or quartet spin-state (e.g., ${}^{4}S-$state). But it is well known that all quartet spin states in ions with three bound 
electrons in the outer-most shells are unstable. In general, such states decay with the emission of one additional electron and formation of another ion with the two 
electrons in its outer-most shells. This simple example illustrates a general experimental situation when some final states in ions arising after the nuclear 
$\beta^{-}-$decay with additional electron ionization, Eq.(\ref{equa2}), are unstable and they can only be stabilized by emitting one additional electron. In other 
the ionization during the nuclear $\beta^{\pm}$-decay can lead, in principle, to very substantial electronic reconfiguration in the final ion. The process, 
Eq.(\ref{eq1}), usually produces a very few minimal changes in the electronic structure of the arising ion in comparison to the incident atom.

Our main goal in this study is the analysis of electron ionization during the nuclear $\beta^{-}$-decay of light atoms and ions. The first actual problem here is related 
to analytical and/or numerical calculations of the overlap integrals which can be found in expressions for the probability amplitudes $M_{if}$ (see below). Some general 
formulas for these overlap integrals are discussed in the next Section. In Section III we discuss the explicit construction of accurate wave functions for few-electron 
atoms/ions and wave function of the free electron moving in the Coulomb field. Actual calculations of the overlap integrals with such wave functions are considered in 
Section IV. Here we develop an original approach which can be used in calculations of the probability amplitudes $M_{if}$ and total probabilities $P_{if}$. This approach 
is based on the use of natural orbital expansions for the incident and final wave functions of atomic systems which take part in the $\beta^{-}$ decay. In the Appendix, 
we discuss the problems related to a restricted accuracy of the sudden approximation which is extensively applied to describe the $\beta^{\pm}$-decay in atoms and 
molecules. 

\section{Final state probabilities}

As is well known the velocities of $\beta^{-}$-electrons ($v_{\beta}$) emitted during the nuclear $\beta^{-}-$decay are significantly larger than usual velocities of atomic 
electrons $v_a$. In particular, in light atoms we have $v_{\beta} \ge 1000 v_a$. This also true for the velocities of the secondary $\delta-$electrons $e^{-}$ which can be 
emitted as `free' particles during the reaction, Eq.(\ref{eq2}), i.e. $v_{\beta} \gg v_{\delta}$. The inequality $v_{\beta} \gg v_a$ allows one to apply the sudden 
approximation and analyze the nuclear $\beta^{-}$-decay in light atoms by calculating the overlaps of the non-relativistic atomic wave functions. The sudden approximation 
assumes that the wavefunction of incident system does not change during the fast process, i.e. its amplitude and phase do not change. By using the sudden approximation we 
can write the following expression for the final state probability of the process, Eq.(\ref{eq1}): 
\begin{equation}
 P_{if} = \mid M_{if} \mid^2 =  \mid \int \Psi_{{\rm Li}}({\bf x}_1, {\bf x}_2, {\bf x}_3) \Psi_{{\rm Be}^{+}}({\bf x}_1, {\bf x}_2, {\bf x}_3) 
 d^3{\bf r}_1 d^3{\bf r}_2 d^3{\bf r}_3 ds_1 ds_2 ds_3 \mid^2 \label{eq3}
\end{equation}
where the notation $M_{if}$ designates the probability amplitude. Here and below the four-dimensional variable ${\bf x}_i$ for $i$-th electron designates combination of the 
three-dimensional position variable ${\bf r}_{i}$ and the spin variable $s_i$, i.e. ${\bf x}_i = ({\bf r}_i, s_i)$ for $i$ = 1, 2, 3. The notations $\Psi_{{\rm Li}}({\bf x}_1, 
{\bf x}_2, {\bf x}_3)$ and $\Psi_{{\rm Be}^{+}}({\bf x}_1, {\bf x}_2, {\bf x}_3)$ stand for the wave functions of the incident Li-atom and Be$^{+}$ ion (both these systems 
contain three bound electrons). The equation, Eq.(\ref{eq3}), means that the probability amplitude $M_{if}$ equals to the overlap of the two bound state wave functions which 
correspond to the two different atomic systems, e.g., the Li atom and Be$^{+}$ ion. The total number of electrons in the incident and final wave functions is exactly the same. 
Therefore, all calculations of the overlap integral(s), Eq.(\ref{eq3}), are relatively simple even in those cases when highly accurate (or truly correlated) wave functions are 
used for both atomic systems. All calculations of the overlap integral were described in our earlier studies (see, e.g., \cite{Talm}, \cite{Our1}) and here we do not want to 
repeat it.
 
In those cases when one atomic electron becomes free during the nuclear $\beta^{\pm}-$decay we can write the following expression for the final state probability of the 
process, Eq.(\ref{eq2}):
\begin{equation}
 P_{if} = \mid M_{if} \mid^2 =  \mid \int \Psi_{{\rm Li}}({\bf x}_1, {\bf x}_2, {\bf x}_3) \Psi_{{\rm Be}^{2+}}({\bf x}_1, {\bf x}_2) \psi_e({\bf x}_3) 
 d^3{\bf r}_1 d^3{\bf r}_2 d^3{\bf r}_3 ds_1 ds_2 ds_3 \mid^2 \label{eq31}
\end{equation}
where the notations $\Psi_{{\rm Li}}({\bf x}_1, {\bf x}_2, {\bf x}_3)$ and $\Psi_{{\rm Be}^{2+}}({\bf x}_1, {\bf x}_2)$ designate the wave functions of the incident Li-atom and 
final Be$^{2+}$-ion, which contains only two bound electrons. The notation $\psi_e({\bf x}_3)$ in Eq.(\ref{eq31}) stands for the wave function of a free electron which moves in 
the central, Coulomb field of the Be$^{2+}$ ion. All these wave functions can be considered as non-relativistic. Also, without loss of generality we shall assume that all bound 
state wave functions arising in our equations for the amplitudes and probabilities have unit norm. Analytical and/or numerical calculation of the overlap integral, Eq.(\ref{eq31}), 
is a significantly more difficult problem than in the case of Eq.(\ref{eq3}), and will be discussed in the next Section.  

As mentioned above the velocity of the emitted $\beta^{\pm}$-electron ($v_{\beta}$) significantly exceeds the usual electron velocities in the both incident and final atoms. It 
follows from here that some conservation laws must be obeyed for any atomic transition during the nuclear $\beta^{\pm}$-decay of few- and many-electron atoms. In particular, the 
angular momentum $L$, spin $S$ and spatial parity $\pi$ of the wave function of the incident atom are always conserved during such a sudden process. For the angular momentum $L$ of 
the final Be$^{2+}$ ion this selection rule allows one to predict some quantum numbers of the final Be$^{2+}$ ion. For instance, if the incident Li atom is in one of its $L = 0$ 
states, then the angular momenta of the final Be$^{2+}$ ion and free electron must be equal to each other. In other words, if the final Be$^{2+}$ ion is formed in the bound $L$-state, 
then the final (or `free') electron is also moving out in an $\ell-$wave, where $\ell = L$. In general, if the incident atom was in the $L_i$-state and final atom is found in one of 
its $L_f$-states, then the final electron is emitted in the $\ell_f$-wave, where $\ell_f$ equals one of the following numbers $\mid L_f - L_i \mid, \ldots, L_f + L_i$. From here one 
can derive a number of useful selection rules which drastically simplify all numerical and analytical computations of the transitions probabilities in Eqs.(\ref{eq3}) - (\ref{eq31}). 
The total number of non-zero few-body integrals which must be evaluated numerically and/or analytically is reduced to a relatively small number. 

\section{Wave functions}

To determine the final state probability $P_{if}$, Eq.(\ref{eq31}), one needs to use the explicit expressions for the wave functions of the incident Li atom, of the final Be$^{2+}$ 
ion, and of the electron which moves in the Coulomb field of the Be$^{2+}$ ion. For the ground $2^2S(L = 0)-$state of the Li atom such a wave function $\Psi$ is written in the 
following general form (see, e.g., \cite{Frolov-Li}, \cite{Lars})
\begin{eqnarray}
 \Psi(\bigl\{ r_{ij} \bigr\})_{L=0} = \psi_{L=0}(A; \bigl\{ r_{ij} \bigr\}) (\alpha \beta \alpha - \beta \alpha \alpha) + \phi_{L=0}(B; \bigl\{ r_{ij} \bigr\}) (2
 \alpha \alpha \beta  - \beta \alpha \alpha - \alpha \beta \alpha) \label{psi}
\end{eqnarray}
where $\psi_{L=0}(A; \bigl\{ r_{ij} \bigr\})$ and $\phi_{L=0}(B; \bigl\{ r_{ij} \bigr\})$ are the two independent radial parts (= spatial parts) of the total wave 
function. The notations $\alpha$ and $\beta$ in Eq.(\ref{psi}) are the one-electron spin-up and spin-down functions, respectively (see, e.g., \cite{Dir}). The 
notations $A$ and $B$ in Eq.(\ref{psi}) mean that the two sets of non-linear parameters associated with the radial functions $\psi$ and $\phi$ can be optimized 
independently. Note that each of the radial basis functions in Eq.(\ref{psi}) explicitly depends upon all six interparticle (or relative) coordinates $r_{12}, r_{13}, 
r_{23}, r_{14}, r_{24}, r_{34}$, where the indexes 1, 2, 3 stand for the three electrons, while index 4 means the nucleus. In modern accurate computations of bound 
states the radial parts of the total wave functions, e.g., $\psi_{L=0}(A; \bigl\{ r_{ij} \bigr\})$ and $\phi_{L=0}(B; \bigl\{ r_{ij} \bigr\})$ in Eq.(\ref{psi}) are 
usually represented by their Hylleraas series, e.g., for the $\psi_{L=0}$-functions
\begin{eqnarray}
 \psi_{L=0}(A; \bigl\{ r_{ij} \bigr\}) = \sum^N_{k=1} C_k r^{n_1(k)}_{23} r^{n_2(k)}_{13} r^{n_3(k)}_{12} r^{m_1(k)}_{14} r^{m_2(k)}_{24}
 r^{m_3(k)}_{34} exp(-\alpha r_{14} -\beta r_{24} -\gamma r_{34}) \label{Hyl}
\end{eqnarray}
where the `parameters' $\alpha, \beta$ and $\gamma$ are varied in computations, but they are the same for all basis functions in the $\psi$-expansion, Eq.(\ref{Hyl}). Three other such 
parameters can be found in the second term from Eq.(\ref{psi}), or $\phi$-expansion. The presence of six non-linear parameters in Eq.(\ref{psi}) increases the overall flexibility of the 
method. However, it is not sufficient to provide very high accuracy for three-electron atoms and ions. Briefly, this means that any accurate trial functions $\Psi(\bigl\{ r_{ij} 
\bigr\})_{L=0}$, Eq.(\ref{psi}), must contain a very large number $N$ of basis functions. 

In order to construct a very efficient variational expansion of the wave function of the three-electron atoms and ion in our earlier work \cite{Frolov-Li} we introduced an advanced set 
of radial basis functions for three-electron bound state calculations. In \cite{Frolov-Li} such a bais set was called the semi-exponential, variational basis set. In general, this 
expansion of the radial function $\psi_{L=0}(A; \bigl\{ r_{ij} \bigr\})$ is written in the form
\begin{eqnarray}
 \psi_{L=0}(A; \bigl\{ r_{ij} \bigr\}) = \sum^N_{k=1} C_k r^{n_1(k)}_{23} r^{n_2(k)}_{13} r^{n_3(k)}_{12} r^{m_1(k)}_{14} r^{m_2(k)}_{24}
 r^{m_3(k)}_{34} exp(-\alpha_{k} r_{14} -\beta_{k} r_{24} -\gamma_{k} r_{34}) \label{semexp}
\end{eqnarray}
where $\alpha_k, \beta_k, \gamma_k$ ($k = 1, 2, \ldots, N$) are the varied non-linear parameters. The presence of a large number of varied non-linear parameters in Eq.(\ref{semexp}) is 
the main and very important difference between the traditional Hylleraas variational expansion, Eq.(\ref{Hyl}), and our variational expansion, Eq.(\ref{semexp}).

The two-electron Be$^{2+}$ ion is the bound atomic system with the two bound electrons which are designated below as particles 1 and 2. The wave functions of such 
systems can accurately be approximated with the use of the exponential variational expansion in relative coordinates $r_{32}, r_{31}$ and $r_{21}$ (see, e.g., 
\cite{Fro98} and references therein). For the bound singlet ${}^1S(L = 0)-$states in the two-electron Be$^{2+}$ ion the exponential variational expansion takes the 
form
\begin{eqnarray}
 \Psi = \frac{1}{\sqrt{2}} \Bigl( 1 + \hat{P}_{12} \Bigr) \sum_{i=1}^{N} C_{i} \exp(-\alpha_{i} r_{24} - \beta_{i} r_{14} - \gamma_{i} r_{12}) 
 (\alpha \beta - \beta \alpha) \label{exp1} 
\end{eqnarray}
which is called the exponential variational expansion in the relative coordinates $r_{24}, r_{14}$ and $r_{21}$. The coefficients $C_i$ are the linear (or variational) 
parameters of the variational expansion, Eq.(\ref{exp1}), while the parameters $\alpha_{i}, \beta_{i}$ and $\gamma_{i}$ are the non-linear (or varied) parameters of 
this expansion. In general, the total energy of the ground $1^1S-$state of the Be$^{2+}$ ion uniformly depends upon the total number of basis functions $N$, Eq.(\ref{exp1}), 
used in calculations. The operator $\hat{P}_{12}$ is the permutation of the two identical particles (electrons). Note also that all relative coordinates $r_{ij}$, i.e. 
$r_{14}, r_{24}, r_{34}, r_{12}, r_{13}$ and $r_{23}$ for the three-electron atoms/ions and $r_{14}, r_{24}$ and $r_{12}$ for the two-electron atoms/ions, are translationally 
and rotationally invariant. 

The last factor $\phi_{e}({\bf r}) = \phi_{e}({\bf r}_{34})$ in the overlap integral, Eq.(\ref{eq31}), represents the unbound (final) electron, which moves in the Coulomb field of the heavy 
Be$^{2+}$ ion. The wave function of such an electron is written in the form $\phi_{e}({\bf r}) = \phi_{kl}(r) Y_{lm}({\bf n})$, where $\phi_{kl}(r)$ is the one-electron radial function, 
while $Y_{lm}({\bf n})$ is the appropriate spherical harmonic and ${\bf n} = \frac{{\bf r}}{r}$ is the unit vector associated with the vector ${\bf r}$. The parameter $k$ is the wave number 
which is uniformly related to the energy of the `free' electron $k = \sqrt{\frac{2 m_e E}{\hbar^2}} = \sqrt{2 E}$ (in atomic units $\hbar = 1, m_e = 1, e = 1$). The explicit formula for the 
radial function $\phi_{kl}(r) = \phi_{kl}(r_{34})$ (in atomic units) is (see, e.g., \cite{LLQ}) 
\begin{equation}
 \phi_{kl}(r) = \frac{C_{kl}}{(2 l + 1)!} (2 Q k r)^{l} \cdot \exp(-\imath Q k r) \cdot {}_1F_1\Bigl(\frac{\imath}{Q k} + l + 1, 2 l + 2, 2 \imath Q k r \Bigr) \label{eq71}
\end{equation}
where $\imath$ is the imaginary unit, ${}_1F_1(a,b;x)$ is the confluent hypergeometric function (see, e.g., \cite{GR}) and $C_{kl}$ is the following factor:
\begin{equation}
 C_{kl} = C_{k0} \cdot  \prod^{l}_{s=1} \sqrt{s^2 + \frac{1}{Q^2 k^2}} = \sqrt{\frac{8 \pi Q k}{1 - \exp(-\frac{2 \pi}{Q k})}} \cdot \Bigl\{ \prod^{l}_{s=1} \sqrt{s^2 + 
 \frac{1}{Q^2 k^2}} \Bigr\} \; \; . \; \; \label{eq72}
\end{equation}
In old papers the $C_{kl}$ factor was called the Coulomb penetration factor. In these two equations the parameter $Q$ is the electric charge of the remaining double-charged 
(positive) Be$^{2+}$ ion. In our case this central atomic cluster is the Be$^{2+}$ ion, which contains two bound electrons, i.e. in all formulas above $Q = 2$ (in atomic 
units). In reality, this parameter can slightly be varied (around 2) to obtain better agreement with the experimental data. Such variations formally represent ionizations 
from different electronic shells of the incident Li atom. For $l = 0$ the second product in the right hand side of Eq.(\ref{eq72}) is reduced to the unity.

With the use of Eq.(7.522.9) from \cite{GR} one finds the following expression for the overlap integral between the auxiliary atomic orbitals $r^n exp(-\gamma_n r)$ and radial 
function $\phi_{kl}(r)$ defined by Eq.(\ref{eq71}):  
\begin{equation}
 I = \frac{C_{kl} (2 k)^{l}}{(2 l + 1)!} \cdot \frac{\Gamma(n+l+3)}{(\gamma_n - \imath k)^{n+l+3}} \cdot {}_2F_1\Bigl(\frac{\imath}{k} + l + 1, n + l + 3; 2 l + 2; 
 - \frac{2 \imath k}{\gamma_n - \imath k} \Bigr) \label{hg1}
\end{equation}
The hypergeometric function in this equation can be transformed to its final form with the use of the formula, Eq.(9.131), from \cite{GR} 
\begin{equation}
 {}_2F_1(a, b; c; z) = (1 - z)^a \cdot {}_2F_1(a, b - c; c; \frac{z}{z - 1} \bigr) \label{hg2} 
\end{equation}
Finally, the formula, Eq.(\ref{hg1}), takes the form
\begin{equation}
 I = \frac{C_{kl} (2 k)^{l}}{(2 l + 1)!} \cdot \frac{(n+l+2)!}{(\gamma_n - \imath k)^{n+l+3}} \cdot \Bigl(\frac{\gamma_n - \imath k}{\gamma_n + \imath k}\Bigr)^{(\frac{\imath}{k} 
 + l + 1)} \cdot {}_2F_1\Bigl(\frac{\imath}{k} + l + 1, l - n - 1; 2 l + 2; \frac{2 \imath k}{\gamma_n + \imath k} \Bigr) \label{hg3}
\end{equation}
As follows from Eq.(\ref{hg3}) the expression for $I$ is reduced to a finite sum (polynomial function), if (and only if) $l = 0$. In this case numerical computations of the corresponding 
hypergeometric functions are simiple and easy. However, for $l \geq 1$ to evaluate the hypergeometric functions at arbitrary, in principle, values of the complex argument one needs to apply 
significantly more sophisticated methods. 

\section{Calculations of the ovelap integrals}

In this Section we discuss actual calculations of the overlap integrals, Eq.(\ref{eq31}). In particular, we analyze the two main difficulties which arise during calculations of the overlap 
integral, Eq.(\ref{eq31}). The first of these difficulties follows from the fact that the total numbers of essential (or internal) variables are different in the incident and final wave 
functions. Indeed, in the incident wave function of the Li-atom one finds six inter-particle coordinates, e.g., three electron-nucleus coordinates $r_{4i}$ ($i$ = 1, 2, 3) and three 
electron-electron coordinates $r_{12}, r_{13}, r_{23}$. In the final wave function we have three electron-nucleus coordinates $r_{4i}$ ($i$ = 1, 2, 3) and only one electron-electron coordinate 
$r_{12}$. This means that the two electron-electron coordinates $r_{13}, r_{23}$ are lost during the sudden transition form the incident to the final state in Eq.(\ref{eq2}). Here we cannot 
discuss all aspects of this interesting problem. Note only that there is an effective method which allows one to avoid problems related with different numbers of the essential variables in the 
incident and final wave functions. This method is based on the natural orbital expansions of all few-electron wave functions which are included in the overlap integral, Eq.(\ref{eq31}). Theory 
of natural orbital expansions was developed in the middle of 1950's by L\"{o}wdin (see discussion and references in \cite{MQW}). Below, we restrict ourselves to a very brief description of this 
theory.  

In this method the radial wave functions are represented in the form of products of the natural orbitals $\chi_{k}(r_{i}) = \chi_{k}(r_{iN})$ (the symbol $N$ stands here for the 
nucleus) which are some simple single-electron functions of one radial variable $r_{iN}$ only. In other words, we are looking for the best approximation of the actual wave function
of $K-$electron atomic system by sets of functions each of which depends upon one radial electron-nucleus coordinate $r_{iN}$ ($i = 1, \ldots, K$) only. In our case for the 
three-electron Li-atom and two-electron Be$^{2+}$ ion we can write 
\begin{eqnarray}
  \Psi_{L=0}(\bigl\{ r_{ij} \bigr\})({\rm Li}) &=& \sum^{N_1}_{n=1} C_n \chi^{(1)}_{n}(r_{1}) \chi^{(2)}_{n}(r_{2}) \chi^{(3)}_{n}(r_{3}) \label{no1} \\
  \Psi_{L=0}(\bigl\{ r_{ij} \bigr\})({\rm Be}^{2+}) &=& \sum^{N_2}_{k=1} B_k \xi^{(1)}_{k}(r_{1}) \xi^{(2)}_{k}(r_{2})  \label{no2} 
\end{eqnarray} 
respectively. Here $\chi_{n}(r_{i})$ and $\xi^{(i)}_{n}(r_{i})$ are the (atomic) natural orbitals constructed for the three-electron Li atom and two-electron Be$^{2+}$ ion (see, e.g., 
\cite{David}, \cite{FrSm}). The coefficients $C_n$ and $B_k$ are the coefficients of the natural orbital expansions for the Li atom and Be$^+$ ion, respectively. In general, these 
coefficients are determined as the solutions (eigenvectors) of some eigenvalue problems. Note that each of these natural orbitals depends only upon the corresponding electron-nucleus 
coordinate $r_{i}$ (or $r_{4i}$ in our notations). They do not include any of the electron-electron (or correlation) coordinates. By using the natural orbital expansions for the 
few-electron wave functions the explicit formula for the overap integral simplifies drastically. With the use of the natural orbital expansions all overlap integrals are always 
represented as the product of three one-dimensional integrals, or as finite sums of such products. Briefly, we can say that application of the natural orbital expansions for 
few-electron atomic wave function allows one to reduce calculations of the overlap integrals to a very simple procedure, e.g., for the process, Eq.(\ref{eq2}), one finds
\begin{eqnarray}
 M_{if} &=& \sum^{N_1}_{n=1} \sum^{N_2}_{k=1} C_n B_k \int_{0}^{+\infty} \chi^{(1)}_{n}(r_{1}) \xi^{(1)}_{k}(r_{1}) r^2_1 dr_1 \int_{0}^{+\infty} \chi^{(2)}_{n}(r_{2}) \xi^{(2)}_{k}(r_{2}) 
 r^2_2 dr_2 \times \nonumber \\ 
 & & \int_{0}^{+\infty} \chi^{(3)}_{n}(r_{3}) \phi_{kl}(r_3) r^2_3 dr_3 \label{amplt} 
\end{eqnarray}
where $\phi_{kl}(r_3)$ are the functions from Eq.(\ref{eq71}). In other words, computations of the overlap integrals are reduced to calculations of one-dimensional integrals and products 
of such integrals, this being the main advantage of our method based on the use of natural orbitals. 

The second general difficulty known in calculation of the overlap integral follows from the fact that the explicit form of the wave functions of the bound and continous spectra in few-electron 
atomic systems are substantially different. This leads to the appearence of integrals (see, e.g., Eq.(\ref{hg3})) which can be evaluated only numerically and with a number of difficulties, 
e.g., slow  convergence of the derived expressions. Furthermore, for some values of parameters the formulas used for such integrals become numerically unstable. This problem can be avoided with 
the use of different systems of basis functions to represent the actual wave functions of the continous spectrum of one-electron Coulomb problem. For instance, one can represent the radial wave 
function from Eq.(\ref{eq71}) as a series of other radial functions which have a relatively simple form and set of overlap integrals between these radial functions and auxiliary orbitals $r^n 
exp(-\gamma_n r)$ mentioned above is written in a simple analytical form. Briefly, this means that it is better to apply different complete sets of basis radial functions to represent the actual 
motion of an unbound electron. In particular, we can use the following set of radial functions $\phi_{kl}(r) = \frac{sin k r}{r} = k j_{0}(k r)$, or $\phi_{kl}(r) = \frac{sin k r}{k r} = j_{0}(k 
r)$, where $k = k_0, 2 k_0, 3 k_0, \ldots$. The `proper' radial functions defined in the previous Section, Eq.(\ref{eq71}), are represented as linear combinations (or Fourier integrals) of these 
`new' basis radial functions and vise versa (see, e.g., \cite{MigK}). 
 
\section{Conclusion}

We have considered the problem of analytical and numerical calculation of the final state probabilities for atomic systems arising during the nuclear $\beta^{\pm}$-decay in few-electron atoms. 
In contrast with our earlier studies here we discuss a possibility to observe `additional' electron ionization during the nuclear $\beta^{-}$-decay in few-electron atoms. We investigate a few 
different approaches which can be applied to calculate the overlap integrals in those cases when one of the final electrons becomes `free' after the $\beta^{\pm}-$decay. These overlap integrals 
are needed to determine the final state probabilities. It is shown that one of the best approaches to calculate such integrals is based on the use of natural orbital expansions for the bound 
state wave functions of the incident and final atomic systems.      

\begin{center}
   Appendix 
\end{center}

The procedure developed in this study is substantially based on the sudden approximation which can be applied to all $\beta^{\pm}$-decaying atomic systems, since the velocity of $\beta^{-}$ 
electron $v_{\beta}$ is significantly faster (in 300 - 10,000 times faster) than the average velocities of atomic electrons $v_{a}$. To the lowest order in the ratio $\tau = \frac{v_a}{v_{\beta}}$ 
we have $\tau = 0$ and all conservation laws mentioned in the main text obey rigorously. However, already in the next order approximation upon $\tau$, i.e. when the ratio $\tau$ is small, but $\tau 
\ne 0$, these conservation laws can only be considered as approximate. In reality, the ratio $\tau$ is very small, e.g., $\tau \approx 1 \cdot 10^{-5} - 1 \cdot 10^{-3}$. This explains the well 
known fact that it is hard to detect any deviation from the `exact' conservation laws. Nevertheless, quite a few deviations from `conservation laws' mentioned in the main text can be found in modern 
experiments with the $\beta^{\pm}$ decaying atoms. Recently, we have developed a general theory of post-sudden approximation for the nuclear $\beta^{-}$-decays in atoms and molecules. However, this
general theory of post-sudden approximations during nuclear processes in atoms and molecules is very complex and cannot be presented here. Instead, let us discuss a few basic experiments which were
proposed earlier to test this theory.  
 
In the first group of experiments we need to detect all possible fast and very fast $\delta-$electrons emitted during the nuclear $\beta^{\pm}$-decay in few-electron atoms/ions, i.e. electrons which 
move with the velocities which are significantly larger than typical atomic velocity $v_a$. Formation of such fast electrons means a substantial transition of the momentum from the $\beta^{-}$ electron
to one of the atomic electrons. The probability of `similar' direct electron excitations is very small for one atomic electron ($\approx$ $1 \cdot 10^{-7}$), but for atoms with $N_e$ electrons such a 
probability is $N_e$ times larger. In any case, possible observation of the fast $\delta-$electron(s) will be a very interesting experimental event which contradicts to the predictions of the theory      
based on the sudden approximation. 

The main goal of the second group of experiments is to find any possible deviation from predictions based on the `rigorous' conservation laws which follow directly from the sudden approximation. For 
instance, consider the $\beta^{-}$-decay of the ${}^{3}$H$^{-}$ ion into the ${}^{3}$He atom \cite{Fro98}. Since the ${}^{3}$H$^{-}$ ion has only one bound state (the ground $1^1S-$state), then the 
final ${}^{3}$He atom can be formed only in one of the $n^{1}S-$states of the ${}^{3}$He atom. Suppose, however, that we have detected the final ${}^{3}$He atom in one of its triplet states, e.g., in 
the $2^3S-$state. The formation of the triplet states in this case means an obvious deviation from the `rigorous' conservation laws which follow from the sudden approximation. Our current expectation to 
detect such `non-conditional' atomic states in experiments is one per $\approx$ 10,000 - 17,000 regular events.    

The last group of possible experiments includes accurate observation and spectral analysis of radiation emitted during the nuclear $\beta^{-}$-decays in few- and many-electron atoms. It is known that any 
fast $\beta^{-}$ electron udergoes an additional acceleration (or deceleration) when it leaves the atomic nucleus (see, discussion of this phenomenon in the Appendix A in \cite{Fro2007}). In actual atoms 
and molecules one finds a small additional acceleration related with the electron-electron interactions. Such an acceleration also leads to the emission of radiation which can be registered in actual 
experiements. For few-electron light atoms the emitted radition which is generated by the interaction between fast $\beta^{-}$-electron and atomic electrons can be registered at the radio and far-infrared 
wavelengths.

\end{document}